\documentclass{WorldScient}
\usepackage{amssymb,graphicx}
\newcommand{\descent}{Newton descent}
\newcommand{\Descent}{Newton Descent}
\newcommand{\costFct}{cost function}

\newcommand{\Loop}{L}
\newcommand{\costF}{F^2}  
\newcommand{\lSpace}{\tilde{x}}                 
\newcommand{\lVeloc}{\tilde{v}}
\newcommand{\pVeloc}{v}		

\newcommand{\rf}[1] {~\cite{#1}}
\newcommand{\refref} [1] {Ref.~\citelow{#1}}

\newcommand{\refrefs}[1] {Refs.~\citelow{#1}}

\newcommand{\refeq}  [1] {Eq.~(\ref{#1})}

\newcommand{\reffig} [1] {Fig.~\ref{#1}}

\newcommand{\refFig} [1] {Fig.~\ref{#1}}

\newcommand{\refsect}[1] {Sec.~\ref{#1}}

\newcommand{\beq}{\begin{equation}}
\newcommand{\continue}{\nonumber \\ }
\newcommand{\nnu}{\nonumber}
\newcommand{\eeq}{\end{equation}}

\newcommand{\ie}{{that is}}		

\newcommand{\jacobianM}{Jacobian matrix}	

\newcommand{\reals}{\mathbb{R}}

\newcommand{\BER}[1]{{\mbox{\footnotesize BER}}} 

\newcommand{\pS}{{\cal M}}			
\newcommand{\Mvar}{{A}}	   	   
\newcommand{\monodromy}{{\bf J}}   

\newcommand\pSpace{x}		

\newcommand\period[1]{{T_{#1}}}			
\newcommand{\cl}[1]{{n_{#1}}}	

\newcommand{\unit}{{\bf 1}}

\newcommand {\id}{{\ \hbox{{\rm 1}\kern-.6em\hbox{\rm 1}}}}

\thicklines
\newlength{\Fsize}   
\newlength{\Fdotsize}
\begin{document}

\title{Turbulent fields and their recurrences}

\author{Predrag Cvitanovi{\'c} and Yueheng Lan}

\address{Center for Nonlinear Science,
	 School of Physics,
	 \\
	 Georgia Institute of Technology, 
	 Atlanta 30332-0430, U.S.A.
	 \\
	 E-mail: predrag.cvitanovic@physics.gatech.edu}
\date{March 10 2003, printed \today}
\maketitle
\abstracts{  
We introduce a new variational method for finding periodic orbits of flows and
spatio-temporally periodic solutions of classical field theories, 
a generalization
of the Newton method to a flow in the space of loops.
The feasibility of the method is demonstrated by its application to
several dynamical systems, including the Kuramoto-Sivashinsky system.
          }

\section{Introduction}
Chaos is the norm for generic
Hamiltonian flows, and for path integrals that implies that 
instead of a few, or countably many extremal configurations, 
classical solutions populate fractal sets of saddles. 
For the path-integral formulation of quantum mechanics
such solutions 
were discovered by Gutzwiller\rf{gutbook} who derived a
trace formula that relates 
a semi-classical approximation of the energy eigenspectrum to
the classical periodic solutions.
While the theory has worked very well in quantum mechanical applications, these ideas remain 
largely unexplored in quantum field theory.

The classical solutions for most strongly nonlinear field theories
are nothing like the harmonic oscillator degrees
of freedom, the electrons and photons of QED; they are 
unstable and highly nontrivial, accessible  only by numerical techniques.
The new aspect, prerequisite to
a semi-classical quantization of strongly nonlinear field theories,\rf{CFTsketch}
is the need to determine a large number of
spatio-temporally periodic solutions for a given
classical field theory. Why periodic? 

The dynamics of strongly nonlinear classical fields is turbulent, not 
``laminar'', and how are we to think about turbulent dynamics?
Hopf\rf{Hopf42} and Spiegel\rf{MS66,BMS71,EAS87} have 
proposed that the turbulence in spatially extended systems
be described in terms of recurrent spatiotemporal patterns.
Pictorially, dynamics drives a given spatially extended system through a
repertoire of unstable patterns; as we watch  
a turbulent system evolve, 
every so often we catch a glimpse of a familiar pattern. 
For any finite  spatial resolution, for a finite time 
the system follows approximately
a pattern belonging to a finite 
alphabet of admissible patterns, and the long term dynamics can be thought
of as a walk through the space of such patterns,
just as chaotic dynamics with a  low dimensional
attractor can be thought of as a succession of nearly periodic (but
unstable) motions. So periodic solutions are needed both to quantify ``turbulence'' in classical field theory, and
as a starting point for the semi-classical quantization of a quantum field theory.

There is a great deal of literature on numerical periodic orbit searches.
Here we take as the starting point Cvitanovi\'c et~al. webbook,\rf{QCcourse}
and in \refsect{s:POserchLowDim} briefly review the Newton-Raphson method 
for low-dimensional flows described by ordinary differential equations
(ODEs),  in order to motivate the {\em \descent} approach that we shall use here, 
and show that it is equivalent to a \costFct\ minimization method.

The problem one faces with high-dimensional flows is that their topology is hard to 
visualize, and that even with a decent starting guess for a point on
a periodic orbit, methods like the Newton-Raphson method are likely to fail.
In \refsect{s:POserchLops} we describe a new method for
finding spatio-temporally periodic solutions of extended, infinite dimensional
systems described by partial differential equations
(PDEs), and in  \refsect{s:ExtMeth} we discuss a simplification
of the method specific to Hamiltonian flows.

The idea is to make an informed rough guess of
what the desired periodic orbit looks like globally, and then use variational methods 
to drive the initial guess toward the exact solution. 
Sacrificing computer memory
for robustness  of the method, we replace a guess that a  {\em point} is on the periodic
orbit by a guess of the {\em entire orbit}. 
And, sacrificing speed for safety, we
replace the Newton-Raphson {\em iteration} by the \descent,
a differential {\em flow} that minimizes 
a \costFct\ computed as deviation of the approximate flow from the true flow
along a smooth loop approximation to a periodic orbit. 

In \refsect{s:ApplicLoopDesc} the method is tested on several systems,
both infinite-dimensional and Hamiltonian, and its virtues, shortcomings 
and future prospects 
are discussed in \refsect{s:SummaryLoopDesc}.

\section{Periodic orbit searches}
\label{s:POserchLowDim}

A periodic orbit is a
solution $(\pSpace,\period{})$, $\pSpace \in \reals^{d}$,
$\period{} \in \reals$ of the {\em periodic orbit condition}
\beq
f^{\period{}}(\pSpace) = \pSpace
\,,\qquad \period{} > 0
\label{e:periodic}
\eeq
for a given flow or mapping $\pSpace \to f^{t}(\pSpace)$.
Our goal here is to determine periodic 
orbits of flows defined by first order ODEs
\beq
\frac{d\pSpace}{dt}=\pVeloc(\pSpace)
	\,,\qquad
 \pSpace \in \pS \subset \mathbb{R}^d
	\,,\;\; 
 (\pSpace,v) \in \bf{T}\pS  
\label{p-1}
\eeq
in many (even infinitely many) dimensions $d$. Here $\pS$ is the phase space 
(or state space) in which evolution takes place, 
$\bf{T}\pS$ is the tangent bundle,\rf{arnold92}
and the vector field $\pVeloc(\pSpace)$ is assumed smooth (sufficiently differentiable).

A {\em prime} cycle
$p$ of period $\period{p}$ is a single traversal of the orbit.
A cycle point of a flow which
crosses a Poincar\'e section $\cl{p}$ times is a fixed point
of the $f^\cl{p}$ iterate of the Poincar\'e section return map $f$,
hence one often refers to a cycle as a ``fixed point''.
By {cyclic invariance}, stability eigenvalues and
the period of the cycle are independent of the
choice of an initial point, so
it suffices to solve \refeq{e:periodic} at a single
cycle point.
Our task is thus to find
a cycle point $\pSpace \in p$ and the
shortest time $\period{p}$ for which \refeq{e:periodic} has a solution.

If the cycle is an attracting limit cycle with a sizable basin of attraction,
it can be found by integrating the flow for sufficiently long time. If
the cycle is unstable, simple integration forward in time will not reveal it,
and methods to be described here need to be deployed.
In essence, any method for solving numerically the periodic orbit
condition 
$F(\pSpace)=\pSpace-f^{\period{}}(\pSpace)=0$
is based on devising a new dynamical system which possesses
the same cycle, but for which this cycle is attractive.
Beyond that, there is a great freedom in constructing such systems, and
many different methods are used in practice.

\subsection{\Descent\ in 1 Dimension}     

\begin{figure}[t] 
\centering
(a)\includegraphics[width=0.5256\textwidth]{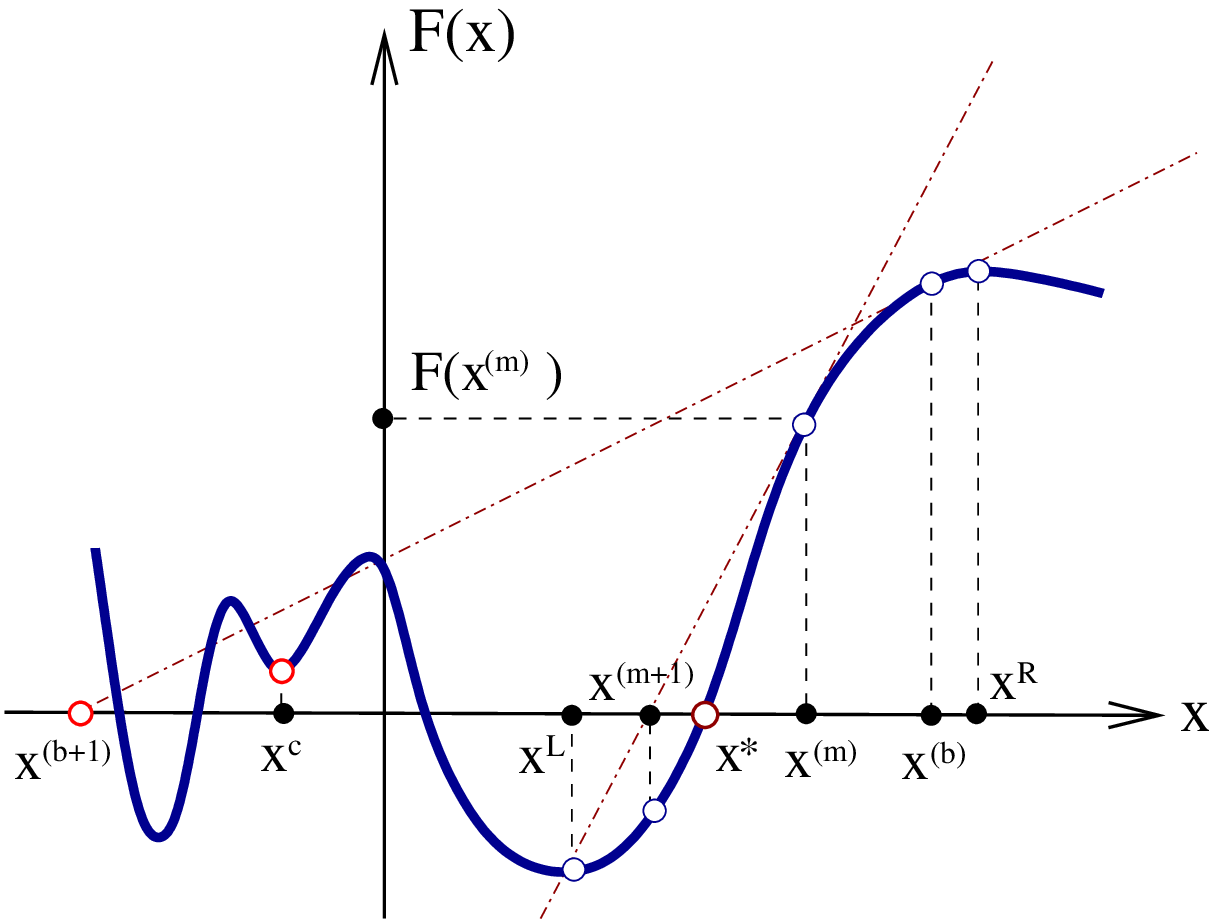}
(b)\includegraphics[width=0.3744\textwidth]{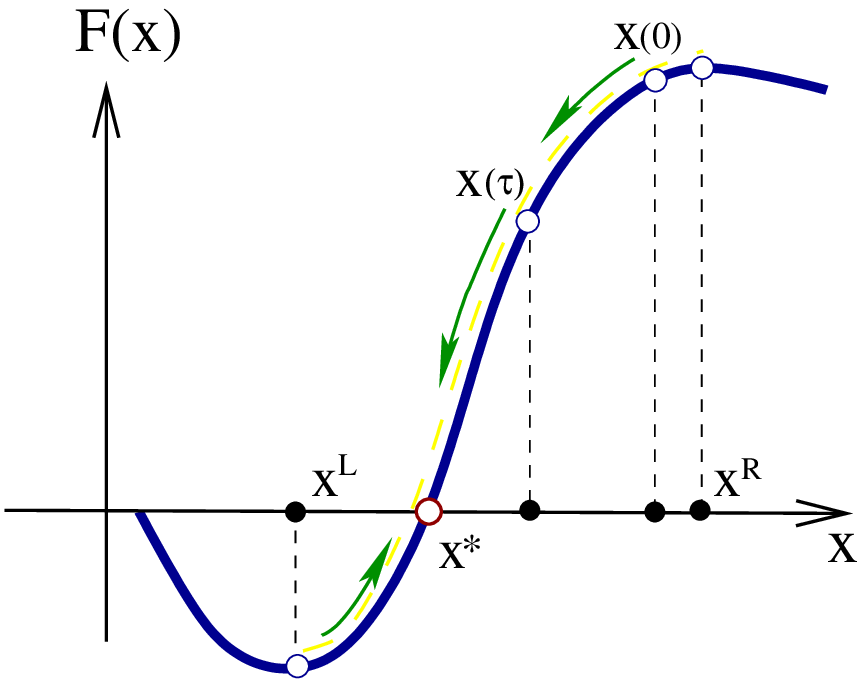}
\caption{
 (a) Newton method: bad initial guess $\pSpace^{(b)}$ 
     leads to the Newton estimate $\pSpace^{(b+1)}$
     far away from the desired zero of $F(\pSpace)$.
     Sequence $\cdots,\,\pSpace^{(m)},\,\pSpace^{(m+1)},\cdots$, 
     starting with a good guess converges super-exponentially
     to $\pSpace^*$.
 (b) \descent: any initial guess $\pSpace(0)$ 
     in the monotone interval $[\pSpace^L,\pSpace^R]$
     of $F(\pSpace)$ flows to $\pSpace^*$ 
     exponentially fast. 
 Both methods diverge if they fall into the basin of attraction of
 a local minimum $\pSpace^{c}$.
        }
\label{f:NewtonDescent}
\end{figure}

Newton's method for determining a zero $\pSpace^*$ of a function $F(\pSpace)$ of one 
variable is based on
a linearization around a starting guess $\pSpace^{(0)}$:
\beq
F(\pSpace) \approx F(\pSpace^{(0)})+F'(\pSpace^{(0)})(\pSpace-\pSpace^{(0)}).
\label{fctayl}
\eeq
An improved approximate solution $\pSpace^{(1)}$ of $F(\pSpace)=0$ is then
$
\pSpace^{(1)} = \pSpace^{(0)} - F(\pSpace^{(0)})/F'(\pSpace^{(0)}) 
\,.
$
{\em Provided} that the $m$th guess 
is sufficiently close to $\pSpace^*$,
the Newton iteration
\beq
\pSpace^{(m+1)}=\pSpace^{(m)}-F(\pSpace^{(m)}) / F^{\prime}(\pSpace^{(m)}) 
\label{c-1}
\eeq     
converges to $\pSpace^*$ super-exponentially fast  (see \reffig{f:NewtonDescent}~(a)). 
In order
to avoid jumping too far from the desired $\pSpace^*$, one often initiates the search by 
the {\em damped Newton method},
\[
\Delta \pSpace^{(m)}=\pSpace^{(m+1)}-\pSpace^{(m)}=
		-{F(\pSpace^{(m)}) \over F^{\prime}(\pSpace^{(m)})}\,\Delta \tau
	\,,\qquad 
0 < \Delta \tau \leq 1
	\,,
\]
takes small $\Delta \tau$ steps at the beginning, reinstating to the full 
$\Delta \tau=1$ jumps only when sufficiently close to the desired $\pSpace^*$.

Let us now take the extremely cautious approach of keeping all steps 
{\em infinitesimally} small, and replacing the discrete sequence $\pSpace^{(m)}, 
\pSpace^{(m+1)}, \ldots $ 
by the fictitious time $\tau$ flow $\pSpace=\pSpace(\tau)$:  
\begin{equation}
d\pSpace=-\frac{F(\pSpace)}{F^{\prime}(\pSpace)} d \tau
	\,,\qquad
\tau \in [0,\infty]
\,. 
\label{c-2}
\end{equation}
If a simple zero, $F^{\prime}(\pSpace^*)\neq 0$, exists
in any given monotone lap of $F(\pSpace)$, it is the attractive fixed point
of the flow \refeq{c-2}  (see \reffig{f:NewtonDescent}~(b)).

While reminiscent of ``gradient descent'' methods,\rf{bl,nr}
this is a flow, rather than an iteration. For lack of established nomenclature
we shall refer to this method of searching for zeros of $F(\pSpace)$ as the {\em \descent},
and now motivate it by re-deriving it from a minimization
principle. Rewriting \refeq{c-2} in terms of a 
``\costFct'' $F(\pSpace)^2$,
\[
d\tau=-\frac{F^{\prime}(\pSpace)}{F(\pSpace)} d\pSpace=
	- \left(\frac{1}{2}\frac{d}{d\pSpace}\ln F(\pSpace)^2\right)d\pSpace,
\]
and integrating,
\[
\tau=-\frac{1}{2}\int_{\pSpace(0)}^{\pSpace}d\pSpace'
		\left(\frac{d}{d\pSpace}\ln F(\pSpace')^2\right)
    =-\frac{1}{2}
\ln \frac{F(\pSpace)^2}{F(\pSpace(0))^2},
\]
we find that the deviation of $F(\pSpace)$ from $F(\pSpace^*)=0$ decays exponentially with the
fictitious time,
\begin{equation}
F(\pSpace(\tau))=F(\pSpace(0)) \, e^{-\tau}
\,,\quad
 \label{c-3}
\end{equation}
with the fixed point
$
\pSpace^* =\lim _{\tau \rightarrow \infty} \pSpace(\tau)
$
reached at exponential rate. In other words, the {\descent}, derived here as an
overcautious version of the damped Newton method, is a flow that 
minimizes the  \costFct\ $F(\pSpace)^2$.

\subsection{Multi-dimensional \Descent}
\label{s:MultDimDescent}

Due to the exponential divergence of nearby trajectories
in chaotic dynamical systems, fixed point searches based on direct
solution of the fixed-point condition
\refeq{e:periodic} as an initial value
problem can be numerically very unstable. Methods that
start with initial guesses $\lSpace_i$ for a number of points along the
cycle are considerably more robust and safer.
Hence we consider next a set of 
periodic orbit conditions
\beq
F_i(\pSpace)=\pSpace_i-f_i(\pSpace)=0
\,,\qquad \pSpace \in \mathbb{R}^{nd}
\label{mltDimZero}
\eeq
where the periodic orbit traverses $n$ 
Poincar\'{e} sections (multipoint shooting method\rf{bl,QCcourse}),
$f_i(\pSpace)$ is the Poincar\'{e} return map from a section 
to the next one, and
the index $i$ runs over $nd$ values, {\ie} $d$ dimensions for each
Poincar\'{e} section crossing.
In this case the expansion \refeq{fctayl} yields the Newton-Raphson iteration
\beq
\lSpace^{(m+1)}_k=
	  \lSpace^{(m)}_k 
	- \left(
		{\unit \over \unit -\monodromy(\lSpace^{(m)})}  
	  \right)_{kl} F_l(\lSpace^{(m)})
\,,\quad
   \monodromy(\pSpace)_{kl} = 
     {{\partial f_k(\pSpace)} \over {\partial \pSpace_l}} 
\,,  
\label{c-5}
\eeq
where
$
    \monodromy(\pSpace)
$\label{jacoBcret}
is the [$d \cl{} \times d \cl{}$] \jacobianM\ of the map $f(\pSpace)$,
and $\lSpace^{(m)}$ is the $m$th Newton-Raphson cycle point estimate.

The \descent\ method \refeq{c-2} now takes form
\beq
\frac{\partial F_i(\lSpace)}{\partial \pSpace_j} d\pSpace_j=-F_i(\lSpace)d \tau
\,. 
\label{c-6}
\eeq
Contracting both sides with $F_i(\lSpace)$ and integrating, we find that
\beq
\costF(\lSpace)=\sum_{i=1}^{d\cl{}} F_i(\lSpace)^2   
\label{c-7}
\eeq
can be interpreted as the \costFct\ \refeq{c-3}, 
also decaying exponentially,
$
\costF(\lSpace(\tau))=\costF(\lSpace(0)) \, e^{-2\tau}
$,
with the fictitious time gradient flow \refeq{c-6} now taking a multi-dimensional form:
\beq
\frac{d\lSpace}{d \tau}=
	\, - \,\frac{\unit}{\unit -\monodromy(\lSpace)} F(\lSpace)
\,.
\label{c-8}
\eeq

Biham and collaborators\rf{biham_wenzel_89}  (see  also  
\refrefs{Schmelcher97,ChandreDasBuch}) were the first to introduce a
fictitous time flow in  searches for periodic orbits of low-dimensional maps, 
with a diagonal matrix with
entries $\pm 1$ in place of the 
$1/({\partial F / \partial\pSpace})$ matrix in \refeq{c-8}. As we lack good
visualizations of high-dimensional flows, for us the ${\partial F / \partial\pSpace}$ matrix 
is essential in determining the direction in which the cycle points should be adjusted.

Here we have considered the case of the guess
$\lSpace$ a vector in a finite-dimensional vector space,
with $\costF(\lSpace)$ the penalty for the distance of $F(\lSpace)$ from its
zero value at a fixed point $\pSpace^*$. 
Our next task is to generalize the {\costFct} 
to a {\em \costFct al} $\,\,\costF[\lSpace]$ which measures the
distance of a loop $\lSpace(s)\in \Loop(\tau)$ from a periodic orbit 
$\pSpace(t) \in p$.

\section{\Descent\ in Loop Space}
\label{s:POserchLops}

For a flow described by a set of 
ODEs, multipoint shooting method of \refsect{s:MultDimDescent} 
can be quite efficient. 
However, multipoint shooting requires 
a set of phase space Poincar\'{e} sections such that an orbit leaving
one section reaches the next one in
a qualitatively predictable manner, without traversing other sections
along the way. In turbulent, high-dimensional
flows such sequences of sections are hard to come by.
One cure for this ill might be a large set of Poincar\'{e} sections, 
with the intervening flight segments short and controllable. Here 
we shall take another path, and discard fixed Poincar\'{e} sections altogether.

    Emboldened by
success of methods such as the multipoint shooting (which eliminates the 
long-time exponential
instability by splitting an orbit into a number of short segments, each with a 
controllable expansion rate) and the cyclist relaxation methods\rf{QCcourse}
(which replace map iteration 
by a contracting flow whose attractor is the desired periodic orbit of the original
iterative dynamics), we now propose a method in which the initial guess is not
a finite set of points, but an entire smooth, differentiable closed loop.

    A general flow \refeq{p-1} has no extremal principle associated with it (we discuss
the simplification of
our method in the case of Hamiltonian mechanics in \refsect{s:ExtMeth}), so there is a 
great deal of arbitrariness in constructing a flow in a loop space. We shall introduce
here the simplest \costFct\ which penalizes mis-orientation of the local loop tangent
vector $\lVeloc(\lSpace)$ relative to the dynamical velocity field $\pVeloc(\lSpace)$ of \refeq{p-1},
and construct a flow in the loop space which minimizes this function. This flow is 
corralled by general topological features of the dynamics,
with rather distant 
initial guesses converging to the desired orbit. Once the loop is sufficiently close
to the periodic orbit, faster numerical algorithms can be employed to pin it down.

\begin{figure}[t] 
\centering
(c) \includegraphics[width=2.5cm]{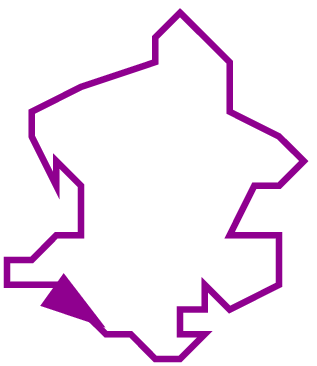}
\hspace{0.1in}
(b) \includegraphics[width=3.5cm]{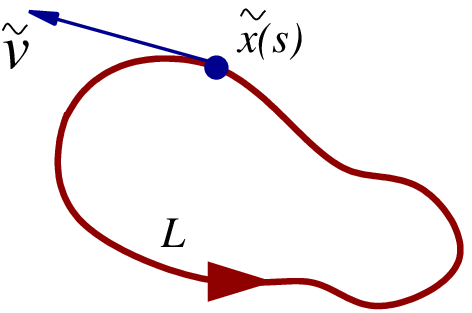}
\hspace{0.1in}
(c) \includegraphics[width=4.0cm]{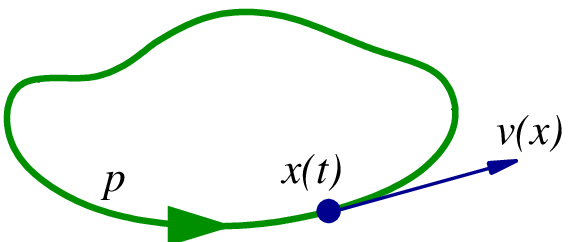}
\caption{
 (a) A continuous path; (b) a loop $\Loop$ with its tangent velocity vector $\lVeloc$;
 (c) a periodic orbit $p$ defined by the vector field $\pVeloc(\pSpace)$.
        }
\label{f:loops}
\end{figure}

    In order to set the notation, we shall distinguish between (see \reffig{f:loops}):

\medskip
\noindent
 {\bf closed path:}
 any closed (not necessarily differentiable) continuous curve 
$J \subset \pS$.

\medskip
\noindent
{\bf loop:}
 a smooth, differentiable closed curve $\lSpace(s)\in \Loop \subset 
\pS$, 
parametrized by $s \in [0,2\pi]$ with $\lSpace(s)=\lSpace(s+2\pi)$, with the
magnitude of the loop tangent vector fixed by 
the (so far arbitrary) parametrization of the loop,
\[
\lVeloc(\lSpace)=\frac{d \lSpace}{ds}\,, \quad \lSpace=\lSpace(s) \in \Loop
\,.
\]   
{\bf annulus:} 
 a smooth, differentiable surface $\lSpace(s,\tau)\in \Loop(\tau)$ swept by a 
family of loops $\Loop(\tau)$, by integration along a fictitious time flow
(see \reffig{f:velocField}~(a))
\[
\dot{\lSpace}=\frac{\partial \lSpace}{\partial \tau}
\,.
\]
{\bf periodic orbit:}
 given a smooth vector field $\pVeloc=\pVeloc(\pSpace),\; (\pSpace,\pVeloc) \in {\bf T} \pS$, periodic orbit $\pSpace(t) \in p$ is a solution of
\[
\frac{dx}{dt}=\pVeloc(\pSpace) 
	\,,\quad
	\mbox{ such that } \pSpace(t)=\pSpace(t+\period{p}),
\] 
where $\period{p}$ is the shortest period of $p$.

\subsection{\Descent\ in the Loop Space}

\begin{figure}[t] 
\centering
(a)\,\,\, \includegraphics[width=4.0cm]{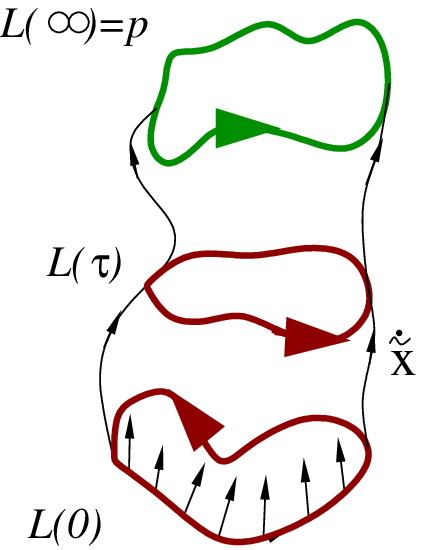}
\hspace{0.1in}
(b) \includegraphics[width=6.5cm]{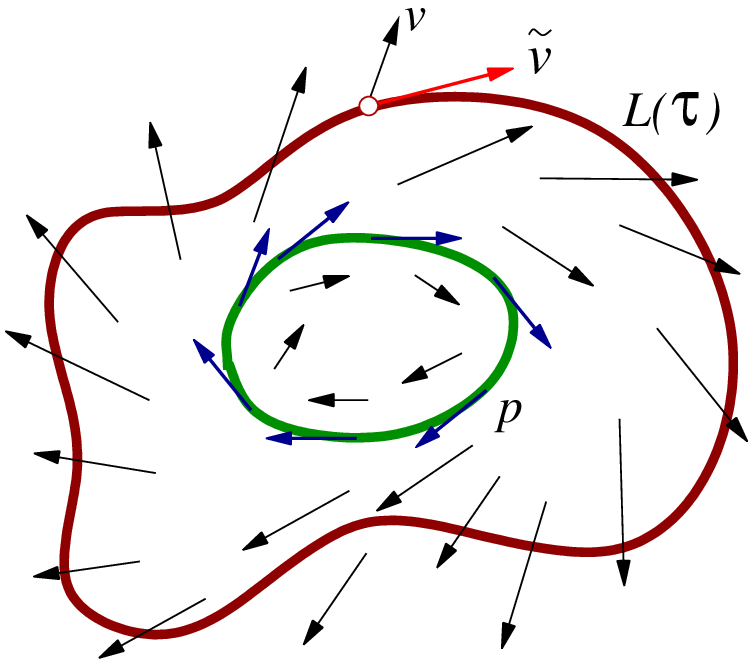}
\caption{
(a) An annulus $\Loop(\tau)$ with vector field $\dot{\lSpace}$ 
connecting smoothly the 
initial loop $\Loop(0)$ to a periodic orbit $p$.
(b) In general the orientation of the loop
tangent $\lVeloc(\lSpace)$ does not coincide with 
the orientation of the velocity field $\pVeloc(\lSpace)$;
for a periodic orbit $p$ it  does so at every $x \in p$.        }
\label{f:velocField}
\end{figure}

In the spirit of \refeq{c-7}, we now define a {\costFct}al for a loop and the 
associated fictitious time $\tau$ flow which sends an initial loop $\Loop(0)$
via a loop family $\Loop(\tau)$ into the periodic orbit $p=\Loop(\infty)$,
see \reffig{f:velocField}~(a).
The only thing that we are given is the velocity field $\pVeloc(\pSpace)$,
and we want to 
 ``comb'' the loop $\Loop(\tau)$ in such a way that 
its tangent field $\lVeloc$ aligns 
with $\pVeloc$ everywhere,
see \reffig{f:velocField}~(b). 
The simplest \costFct al for the job is 
\beq
\costF[\lSpace] = {1 \over 2\pi}
	        \oint_\Loop ds\,(\lVeloc-\lambda\, \pVeloc)^2
	\,,\quad
	\lVeloc = \lVeloc(\lSpace(s,\tau))\,,\,\,\,
	\pVeloc = \pVeloc(\lSpace(s,\tau))	   
\,.
\label{loopCostFct}
\eeq
As we have fixed the loop period to $s=2\pi$, the parameter 
$\lambda = \lambda(s,\tau)$ is 
needed to match the magnitude of the tangent field
$\lVeloc$ (measured in the loop
parametrization units $s$) to the velocity field $\pVeloc$ 
(measured in the dynamical time units $t$).
$|\pVeloc|$ cannot vanish anywhere along the loop  $\Loop(\tau)$,
as in that case the loop would be passing through an equilibrium point,
and have infinite period.
We shall take a very simple  choice, and set $\lambda$
to be 
a global, space-independent loop parameter $\lambda = \lambda(\tau)$.
In the limit where the loop is the desired periodic orbit $p$,
this $\lambda$ is
the ratio of the dynamical period $\period{p}$ 
to the loop parametrization period $2\pi$,  $\lambda={\period{p}}/{2\pi}$.
More general choices of the parametrization $s$ will be
discussed elsewhere.\rf{LC03}
 
Take a derivative of the  \costFct al $\costF[\lSpace]$
 with respect to the (yet undetermined) fictitious time $\tau$,
\[ 
{d \costF \over d\tau} = {1 \over \pi} \oint_\Loop ds\,
	(\lVeloc-\lambda\, \pVeloc)\,
	{d~ \over d\tau}
	(\lVeloc-\lambda\, \pVeloc)
\,.
\] 
The simplest, exponentially  decreasing 
 \costFct al 
is obtained by taking the $\lSpace(s,\tau)$ dependence on
$\tau$ to be point-wise proportional to the deviation of the
two vector fields
\beq
\frac{d}{d \tau}(\lVeloc-\lambda \pVeloc)=
 -(\lVeloc-\lambda \pVeloc)
\,, 
\label{fm0}
\eeq
so the fictitious time flow drives the
loop to $\Loop(\infty)=p$, see \reffig{f:velocField}~(a):
\beq
\lVeloc-\lambda \pVeloc=   e^{-\tau} (\lVeloc- \lambda \pVeloc)|_{\tau=0}
\,. 
\label{fm}
\eeq

Making the $\lSpace$ dependence in \refeq{fm0} explicit
we obtain our main result, the  \descent\ PDE which evolves the initial
loop $\Loop(0)$ into the desired periodic orbit $p$
\beq
\frac{\partial^2 \lSpace}{\partial s \partial \tau}-\lambda \Mvar \frac
{\partial \lSpace}{\partial \tau}-\frac{\partial \lambda}{\partial \tau}\pVeloc
	=
		\lambda \pVeloc-\lVeloc
\,,\qquad
 \Mvar_{ij}(x)=\frac{\partial \pVeloc_i(x)}{\partial x_j}
\label{bd5}
\eeq     
in the fictitious time $\tau \to \infty$.
Here $\Mvar$
is the matrix of variations of the flow (its integral around $p$ yields the linearized stability
matrix for the periodic orbit $p$).

\subsection{Loop Initialization and Numerical Integration}
\label{s:LoopInit}

    Replacement of a finite number of points along a trajectory by a closed smooth 
loop, and of the Newton-Raphson iteration by the \descent\ flow
results in a second order PDE 
for the loop evolution. 
The loop parameter $s$ converges (up to a proportionality constant)
to the dynamical time $t$ as the loop converges to the 
desired periodic orbit. The flow
parameter $\tau$ plays the role of a fictitious time. 
Our aim is 
to apply this method to high-dimensional flows; and thus we have replaced the 
initial ODE dynamics \refeq{p-1} by a very high-dimensional PDE.
And here our troubles start - can this be implemented at all? 
How do we get started?

    A qualitative understanding of the dynamics is the essential prerequisite
to successful periodic orbit searches. We start by long-time numerical runs 
of the dynamics, in order to get a feeling for frequently visited regions of the
phase space (``natural measure''),
and to search for close recurrences. 
 We construct the initial loop $\Loop(0)$ using the intuition so acquired. Taking
a fast Fourier transform of the guess, keeping the lowest frequency 
components, and  transforming back to the initial phase space helps smooth the
initial loop $\Loop(0)$. 
A simple linear stability analysis shows that the smoothness
of the loop is maintained by  the flow in the fictitious time $\tau$.
This, as well as worries about the marginal stability eigenvalues 
and other details of the numerical integration of the
loop flow \refeq{bd5}, are described in 
\refref{LC03}. Suffice it to say that after a considerable amount of computation
one is rewarded 
by periodic orbits that could not have been obtained by the methods employed
previously.

\section{Extensions of the Method}          
\label{s:ExtMeth}

In classical mechanics
particle trajectories are solutions of a 
different variational principle, 
the Hamilton's variational principle.
For example, one can determine a periodic orbit of a 
billiard by wrapping around a rubber band of roughly correct topology, and then
moving the points along the billiard walls until the length ({\ie}, the action) of the
rubber band is extremal (maximal or minimal under infinitesimal changes of the
boundary points). In this case, 
extremization of action requires only 
$D$-dimensional ($D = $ degrees of freedom) rather than $2D$-dimensional 
($2D = $ dimension of the phase space) variations.

Can we exploit this fact to simplify our calculations in Newtonian mechanics?
The answer is yes, and easiest to understand in terms of the Hamilton's
variational principle which states that classical trajectories are extrema
of the Hamilton's principal function (or, for fixed energy, the action)
\[
R(q_1,t_1;q_0,t_0) = \int_{t_0}^{t_1} \! dt \,{\cal L}(q(t), {\dot q}(t),t)
\,,
\]
where
$ {\cal L}(q, {\dot q},t)$ is the Lagrangian.
Given a loop $\Loop(\tau)$ we can compute
not only the tangent ``velocity'' vector $\lVeloc$, but also 
the local loop ``acceleration'' vector 
\[
\tilde{a} = 
\frac{d^2 \lSpace}{d s^2}
\,,
\]
and indeed, as many $s$ derivatives as needed. 
Matching the dynamical acceleration $a(\lSpace)$ with the loop ``acceleration'' 
$\tilde{a}(\lSpace)$
results in an equation for the evolution of the loop
\[
\frac{d}{d \tau}(\tilde{a}-\lambda^2 a)=-(\tilde{a}-\lambda^2 a)
\,,
\]   
where $\lambda^2$ appears instead of $\lambda$ for dimensional reasons.
This equation can be re-expressed in terms of loop variables $\lSpace(s)$;
the resulting equation is somewhat more complicated than \refeq{bd5}, but the saving is significant -
only 1/2 of the phase-space variables appears in the fictitious time flow.
More generally, the method works for extremization of functions of form
$ {\cal L}(q, {\dot q},{\ddot q},\dots,t)$, with considerable computational savings.\rf{LC03}

\section{Applications}
\label{s:ApplicLoopDesc}

We now offer several examples of the application of the 
\descent\ in the loop space, \refeq{bd5}.

\subsection{Unstable Recurrent Patterns in a Classical Field Theory}

One of the simplest and extensively studied
spatially extended dynamical systems is 
the Kuramoto-Sivashinsky system\rf{KurSiv}
\beq
u_t=(u^2)_x-u_{xx}-\nu u_{xxxx} 
\label{ks1}
\eeq
which arises as an amplitude equation for interfacial instabilities
in a variety of contexts.
The ``flame front'' $u(x,t)$ has compact support, with
$x \in [0,2\pi]$ a periodic space coordinate.
The $u^2$ term makes this a nonlinear system,
$t$
is the time,  and $\nu$ is a fourth-order ``viscosity'' damping parameter
that irons out any sharp features. Numerical simulations demonstrate that as
the viscosity decreases (or the size of the system increases),
the ``flame front'' becomes increasingly unstable and turbulent.\rf{KNS90,Foias88}
The task of the theory is to describe this spatio-temporal
turbulence and yield quantitative predictions for its measurable
consequences.

As was argued in \refref{CCP97}, turbulent dynamics of such systems
can be visualized as a walk through the space of unstable spatio-temporally 
recurrent patterns. In the
PDE case we can think of a spatio-temporally discretized guess solution as 
a surface covered with small but misaligned tiles. Decreasing 
\refeq{loopCostFct} by the \descent\ means smoothing these
strutting fish scales into
a smooth surface, a solution of the PDE in question. 

In case at hand it is more convenient to transform the problem to Fourier space.
If we impose the periodic boundary condition $u(t,x+2\pi)=u(t,x)$ and choose
to study only the odd solutions $u(-x,t)=-u(x,t)$,\rf{CCP97}
the spatial Fourier series for the wavefront is 
\begin{equation}
u(x,t)=i\sum_{k=-\infty}^{\infty} a_k(t) \exp (ikt)
\,, 
\label{expan}
\end{equation} 
with real Fourier coefficients $a_{-k}=-a_k$, and \refeq{ks1} takes form
\begin{equation}
\dot{a_k}=(k^2-\nu k^4)a_k-k\sum_{m=-\infty}^{\infty}a_m a_{k-m} \,.
\label{ksf}
\end{equation}
After the initial transients die out,
for large $k$ the magnitude of $a_k$ Fourier component
decreases exponentially with  $k^4$, justifying use of Galerkin truncations 
in numerical simulations. As in numerical work on any PDE
we thus replace \refeq{ks1} 
by a finite but high-dimensional system of ODEs. 
The initial searches 
for the unstable recurrent patterns 
for this spatially extended system found several hundreds of periodic
solutions close to the onset of spatiotemporal
chaos,
but a systematic exploration of
more turbulent regimes was unattainable by the
numerical techniques employed.\rf{CCP97,ZG96}

\begin{figure}[t] 
\centering
(a) \includegraphics[width=2.0in]{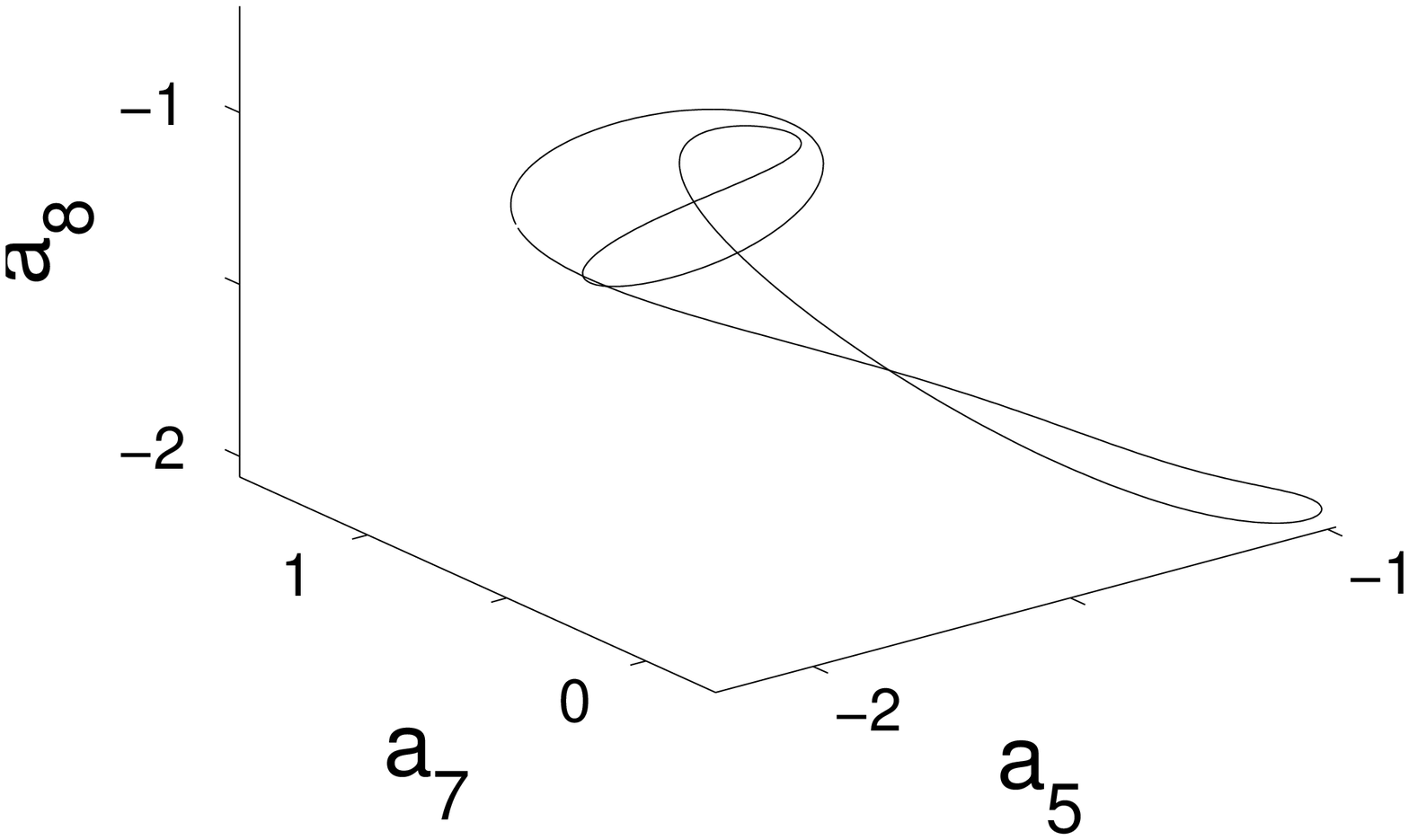}
\hfill
(b) \includegraphics[width=2.0in]{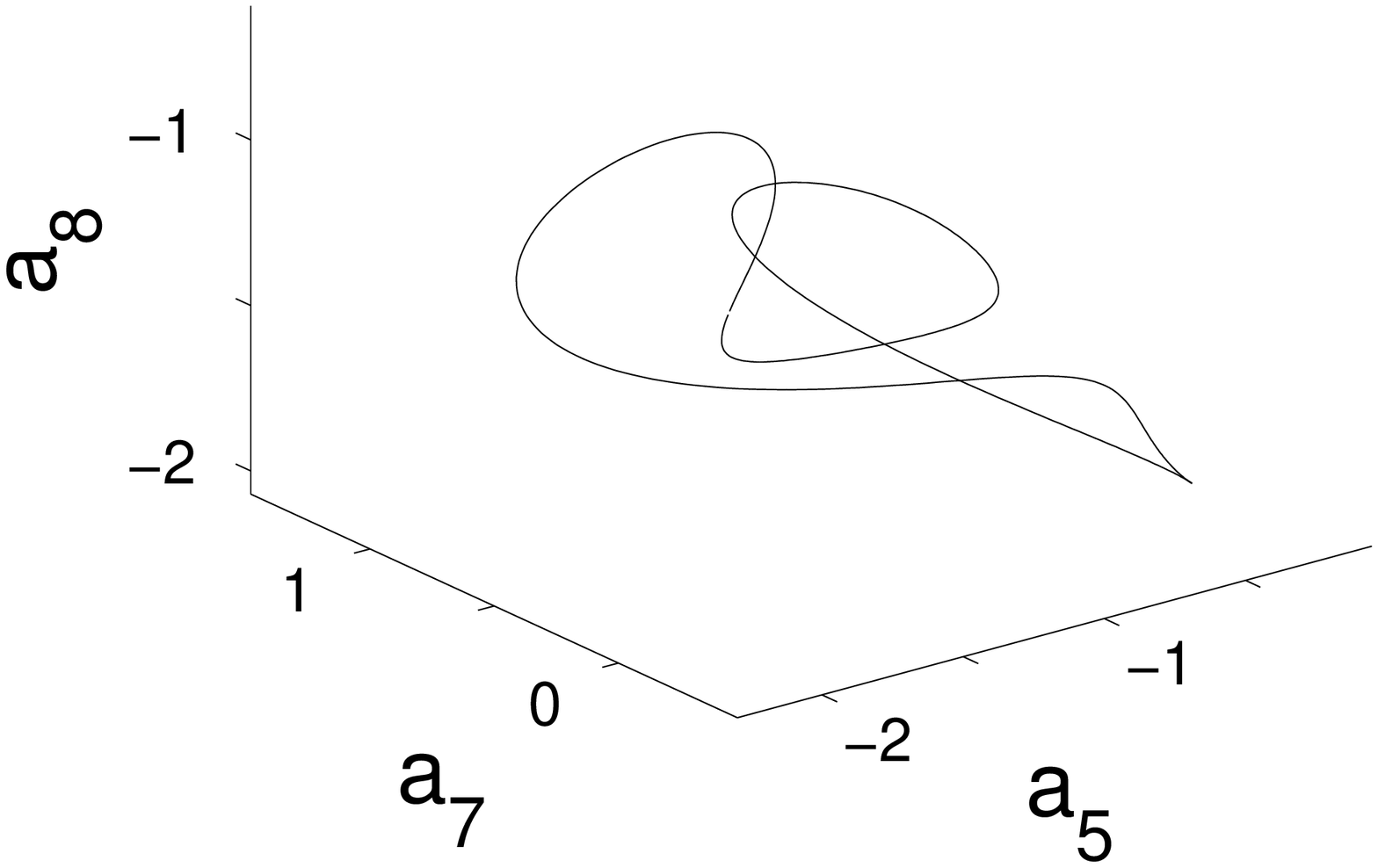}

\caption{
 (a) An initial guess $\Loop(0)$, and 
 (b) the periodic orbit $p$ of period $\period{p}= 0.5051$ reached 
 by the \descent, the
 Kuramoto-Sivashinsky system in a spatio-temporally
 turbulent regime (viscosity parameter
 $\nu=0.01500$, $d=32$ Fourier modes truncation). 
 In discretization of the initial loop $\Loop(0)$ each point has to be specified in
 all $d$~dimensions;
 here the coordinates $\{a_5,a_7,a_8\}$ are picked arbitrarily,
 other projections from $d=32$ dimensions to a subset of 3 coordinates are
 equally (un)informative.
	 }
\label{f:ks1}
\end{figure}  

With decreasing viscosity $\nu$
the system becomes quite turbulent, with the spatiotemporal portraits
of the flame front $u(x,t)$ a complex labyrinth of eddies of different
scales and orientations, and its Fourier space dynamics \refeq{ksf} a complicated
high-dimensional trajectory.

In \reffig{f:ks1} 
we give an example of a \descent\ calculation for this system,
for the viscosity parameter
$\nu$ significantly lower than in the 
earlier investigations. 
Although the  initial guess $\Loop(0)$ is quite far from the 
final configuration $p=\Loop(\infty)$, the method succeeds in 
molding the starting loop into a periodic solution 
of this high dimensional flow.
A systematic exploration of the shortest cycles
found,  and  the hierarchy of longer
cycles will be reported elsewhere.\rf{LanThesis}

\subsection{H\'enon-Heiles and Restricted 3-body Problems}

Next we offer two examples of the applicability of the extension of 
the \descent\ of \refsect{s:ExtMeth} 
to low-dimensional Hamiltonian flows.

\begin{figure}[t] 
\centering
(a) \includegraphics[width=2.0in]{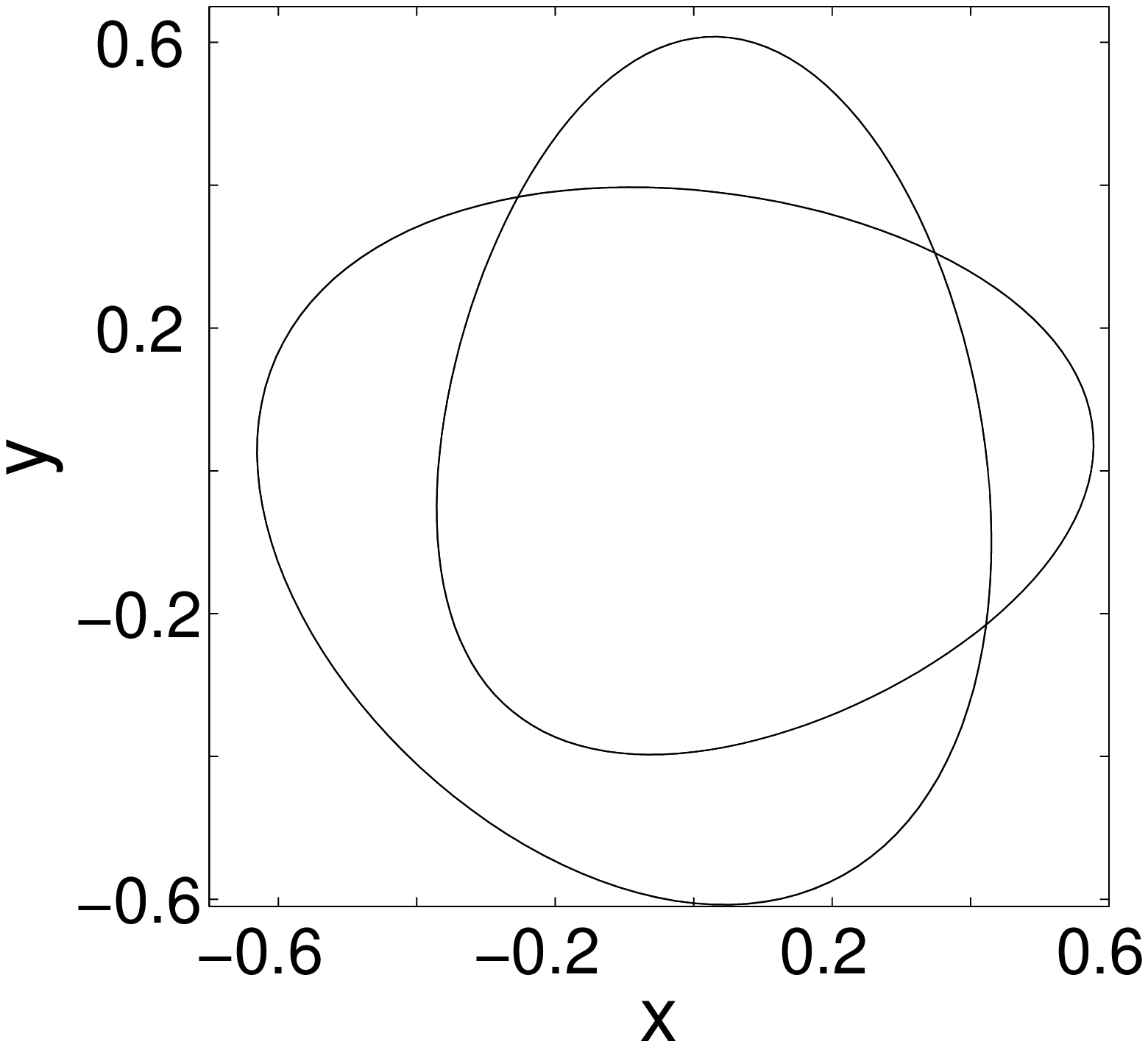}
\hfill
(b) \includegraphics[width=2.0in]{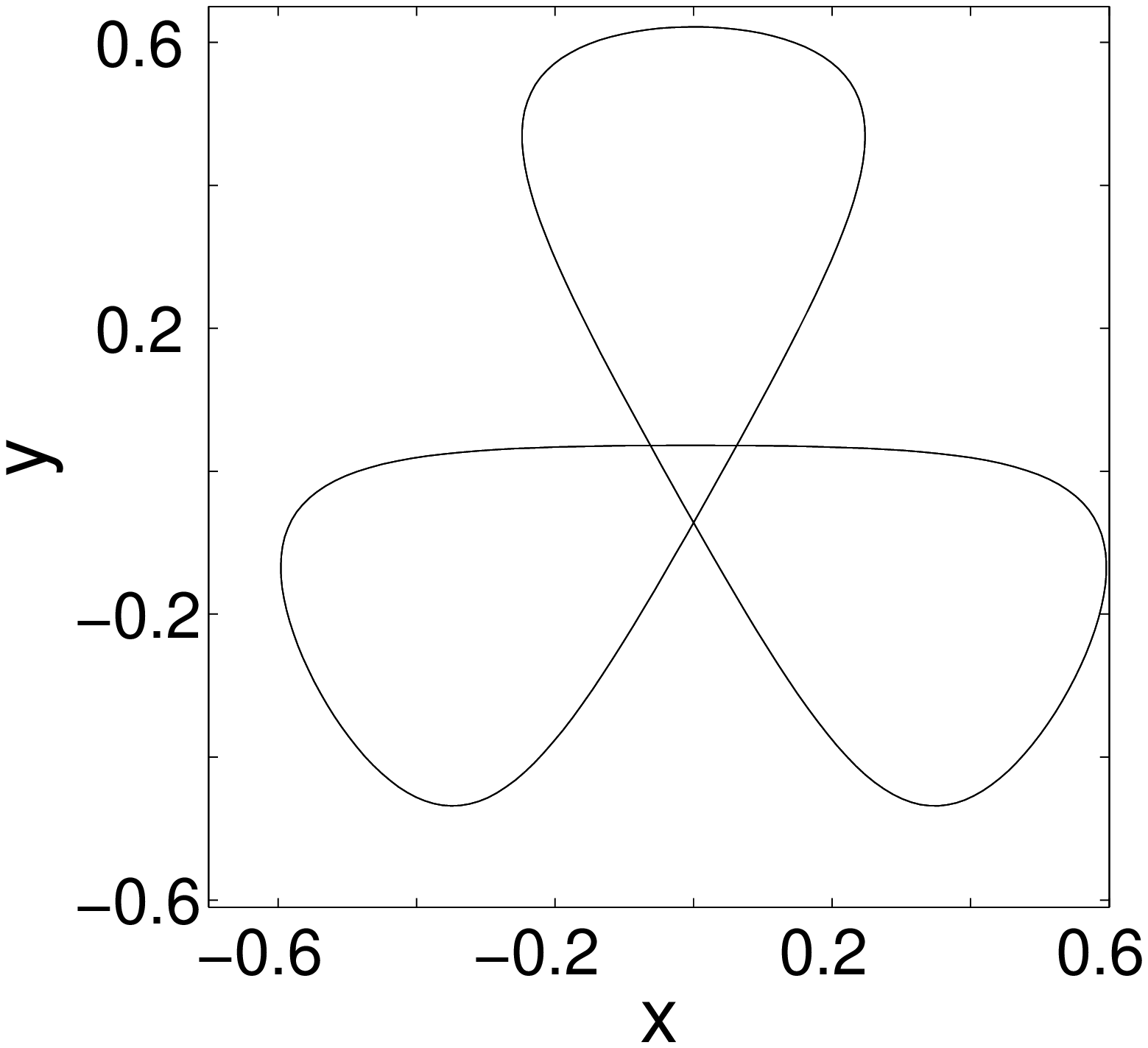}
\caption{
 (a) An initial loop $\Loop(0)$, and 
 (b) the periodic orbit $p$ reached by the \descent, the
 H\'enon-Heiles system in a chaotic regime, $E=0.1794$.
 The period was fixed arbitrarily to $\period{}=13.1947$
 by taking a fixed value of the scaling $\lambda=2.1$.
	 }
\label{f:hh}
\end{figure}

H\'enon-Heiles Hamiltonian\rf{HenHeil}
\beq
H=\frac{1}{2}(\dot{x}^2+\dot{y}^2+x^2+y^2)+x^2 y-\frac{y^3}{3} 
\label{hheq}
\eeq
is frequently used in astrophysics.
\refFig{f:hh} shows an
application of the method of \refsect{s:ExtMeth}
to a periodic orbit search restricted
to the configuration space.

\begin{figure}[t] 
\centering
(a) \includegraphics[width=2.0in]{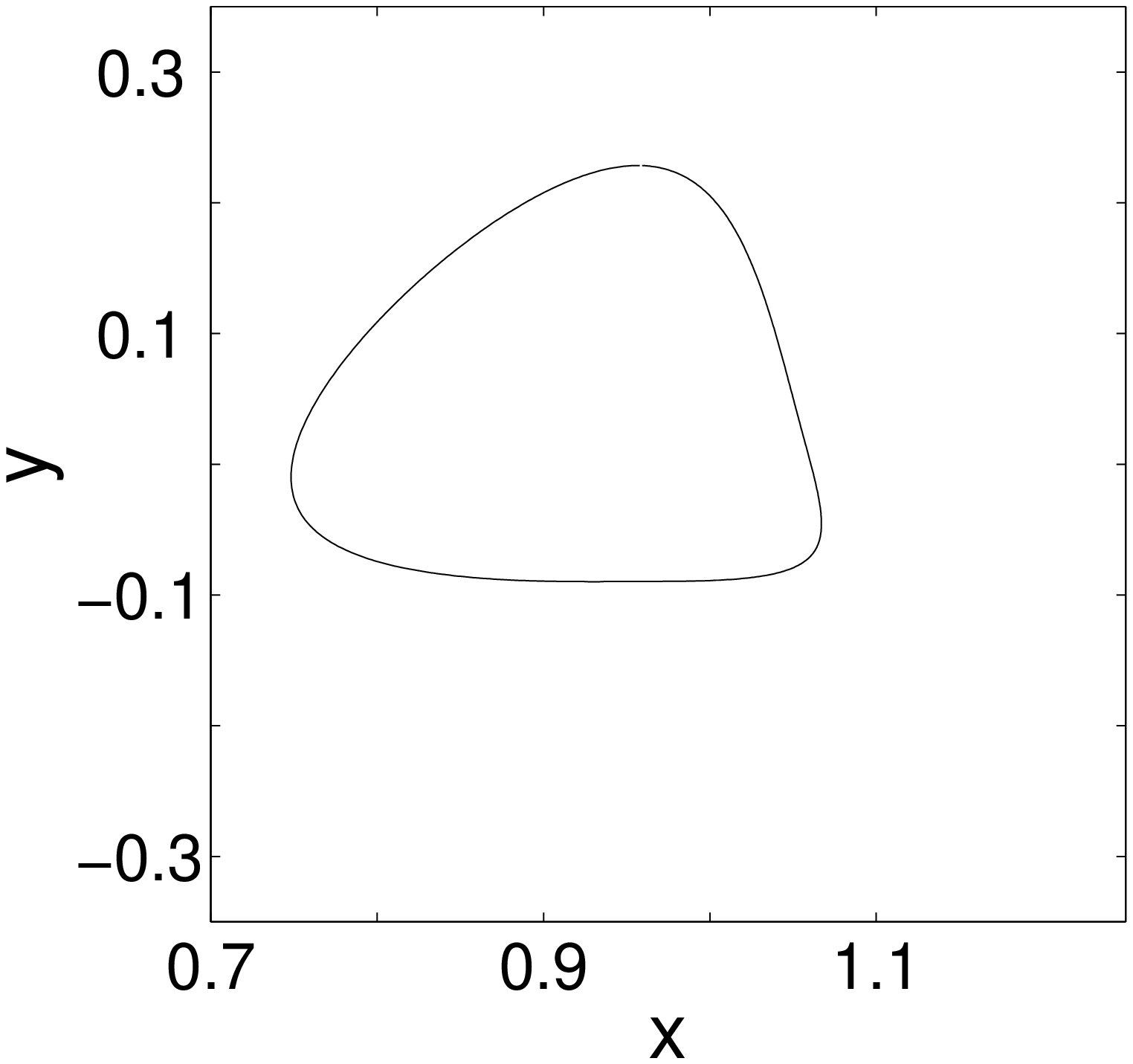}
\hfill
(b) \includegraphics[width=2.0in]{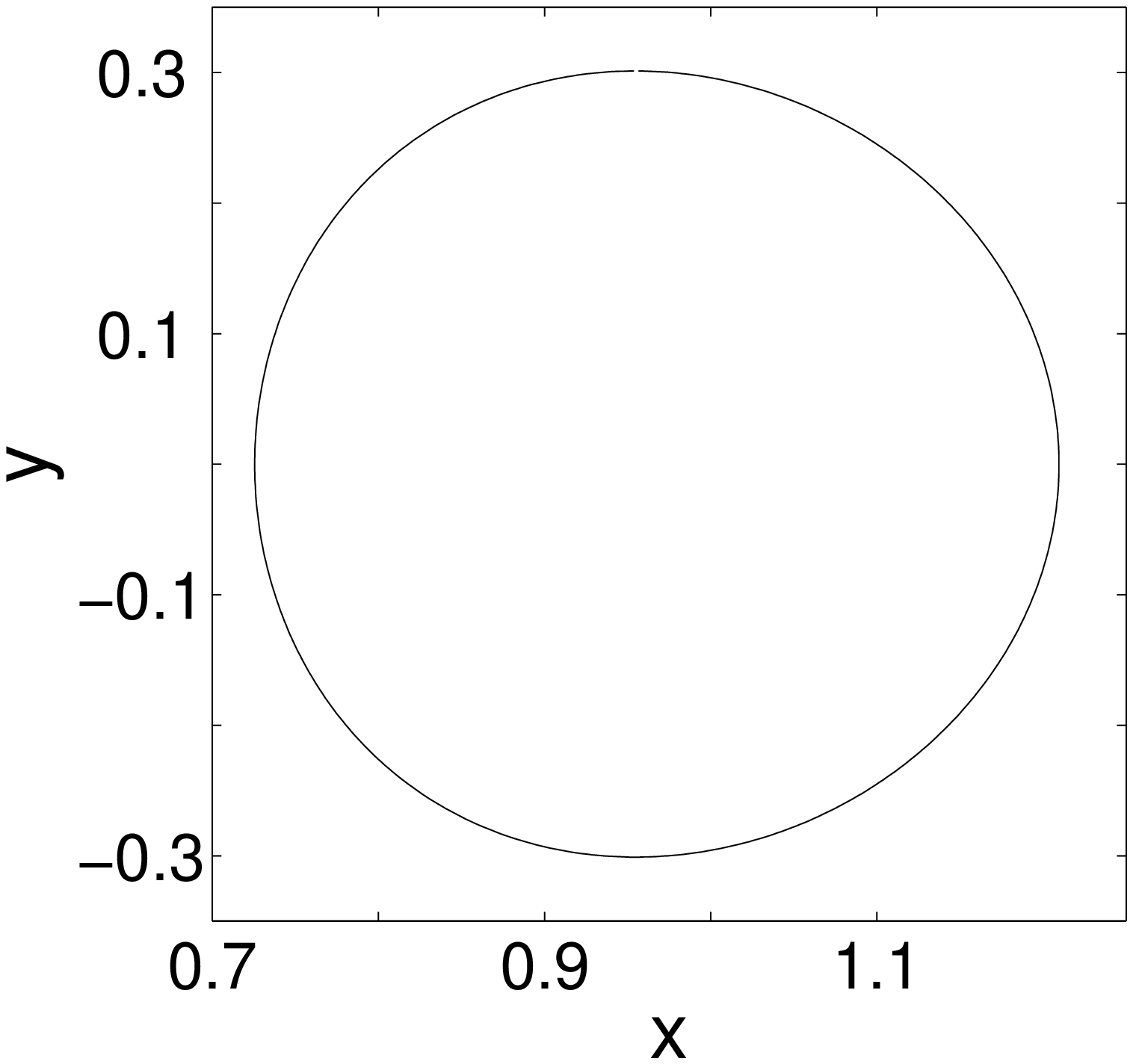}
\caption{
 (a) An initial loop $\Loop(0)$, and 
 (b) the periodic orbit $p$ reached by the \descent, the restricted
 three body problem in the chaotic regime, $\mu=0.04$, $T_p=2.7365$.
	 }
\label{rtb}
\end{figure}

In the H\'enon-Heiles case the acceleration $(a_x,a_y)$ depends
only on the configuration coordinates
$(x,y)$. More generally, the $a$'s could also depend on $(\dot{x},\dot{y})$,
as is the case for 
the restricted three-body problem 
equations of motion\rf{rtbRefs}
\begin{eqnarray}
\ddot{x} &=& 2\dot{y}+x-(1-\mu)\frac{x+\mu}{r_1^3}-\mu \frac{x-1+\mu}{r_2^3} 
	\continue
\ddot{y} &=& -2\dot{x}+y-(1-\mu)\frac{y}{r_1^3}-\mu \frac{y}{r_2^3}
	\label{rtbeq}\\
r_1 &=&  \sqrt{(x+\mu)^2+y^2}
	\,,\qquad
r_2 \,=\,\sqrt{(x-1+\mu)^2+y^2}
\nnu
\end{eqnarray}
which describe the motion of a
``test particle'' in a rotating frame under the influence of the gravitational force of two
heavy bodies with masses $1$ and $\mu \ll 1$, fixed at $(-\mu,0)$ and $(1-\mu,0)$ in 
the $(x,y)$ coordinate frame.
The periodic solutions of \refeq{rtbeq} correspond to periodic or 
quasi-periodic motion of the test particle in the inertial frame. \refFig{rtb} shows 
an application of the \descent\ method to this problem.

\section{Summary and Future Directions}
\label{s:SummaryLoopDesc}

The periodic orbit theory approach to classically turbulent
field theory is to visualize 
turbulence as a sequence of near recurrences in a repertoire of unstable 
spatio-temporal patterns. 
So far, existence of a 
hierarchy of spatio-temporally periodic solutions, and 
applicability of the periodic orbit theory in
evaluation of global averages 
for spatially extended nonlinear system has been demonstrated
in one example, the Kuramoto-Sivashinsky system.\rf{KurSiv}
The parameter ranges previously explored probe
the weakest nontrivial ``turbulence'', and it is an open 
question to what extent the approach remains implementable 
as such classical fields go more turbulent. 
Kawahara
and Kida\rf{KawKida01} were able to find two periodic solutions in a turbulent plane Couette flow,
in a {\em 15,422-dimensional} discretization of the full 3-$d$ Navier-Stokes equations, 
so we remain optimistic.

The bottleneck has been the lack of methods for finding even the simplest
periodic orbits in  high-dimensional flows, and the lack of intuition as
to what such orbits would look like.
Here we have formulated the ``\descent'' method, a very conservative 
method which emphasizes
topological robustness at a considerable cost to numerical speed, 
and demonstrated that the method enables us
to find the shortest
spatio-temporally unstable periodic solutions of an
(infinite dimensional) classical field theory,
as well as several Hamiltonian flows.

In devising the \descent\ method we have made a series of restrictive choices, 
many of which could be profitably relaxed.

The choice of a {\bf Euclidean metric \costFct} $\costF[\lSpace]$ has no compelling
merit, other than notational simplicity. For a flow like the Kuramoto-Sivashinsky
the $a_1$, $a_2$, $\dots$ directions are clearly more important than
$a_k$, $a_{k+1}$, $\dots\;$,  $k$ large, and that is not 
accounted for by the current form of the
\costFct. A more inspired choice would use intrinsic information
about dynamics, replacing $\delta_{ij}F_i F_j$ by 
a more appropriate metric $g_{ij}F_i F_j$ 
that penalizes straying away in the unstable directions
more than deviations in 
the strongly contracting ones.

{\bf Loop parametrization}. 
Once it is understood that given a vector field $\pVeloc(\pSpace)$, the objective is
to determine a loop $\Loop(\infty)=p$
whose tangent vectors point along $\pVeloc(\pSpace)$ everywhere
along the loop, there is no reason to fix  the loop parameter 
$s$ by making it proportional to the dynamical system time $t$.
Any loop parametrization $s$ will do, and other choice might be more 
effective for numerical discretizations.

{\bf Zero modes}.
In numerical calculations we eliminate
the marginal eigendirection along the loop by ``gauge fixing'', 
fixing one point on the loop by an arbitrary Poincar\'{e} section.
This seems superfluous
and perhaps
should be eliminated in favor of
some other, more invariant criterion.

The {\bf  \descent\ method} introduced here replaces the
Newton-Raphson iteration by an exponentially contracting 
flow.
Keeping the fictitious time step $d \tau$ infinitesimal is both against the 
spirit of the Newton method, and not what we do in practice; once the approximate 
loop is sufficiently close to the desired periodic orbit, we
replace $d \tau$ 
by discrete steps of increasing size $d \tau \rightarrow 1$, in order 
to regain the super-exponential
convergence of the Newton method.
    
{\bf Topology}.
 As for high-dimensional flows we are usually clueless as to what the solutions should look like, currently we have no way of telling to which periodic orbit the loop space flow
\refeq{bd5} will take our initial guess, other than to the ``nearest'' periodic orbit of topology 
``similar'' to the initial loop.

\end{document}